\documentclass[]{raa}            
\usepackage{graphicx,times}
\usepackage{natbib}

\providecommand{\bm}[1]{\mbox{\boldmath$#1$\unboldmath}}

\newcommand{\heii}{He \scriptsize{II} \normalsize}
\newcommand{\fexvi}{Fe \scriptsize{XVI} \normalsize}

\begin{document}
\title{An investigation of flare emissions at multiple wavelengths
\footnotetext{\small $*$ Supported by the National Natural Science Foundation of China.}
}

\volnopage{ {\bf --} Vol.\ {\bf --} No. {\bf XX}, 000--000}
   \setcounter{page}{1}

\author{Dong Li\inst{1,2,3}\thanks{Corresponding author}, Alexander~Warmuth\inst{4}, Lei~Lu\inst{1}, and Zongjun~Ning\inst{1,5}}

\institute{$^1$Key Laboratory for Dark Matter and Space Science, Purple Mountain Observatory, CAS, Nanjing 210033, China; {\it lidong@pmo.ac.cn} \\
$^2$CAS Key Laboratory of Solar Activity, National Astronomical Observatories, Beijing 100101, China \\
$^3$State Key Laboratory of Space Weather, Chinese Academy of Sciences, Beijing 100190, China \\
$^4$Leibniz-Institut f\"{u}r Astrophysik Potsdam (AIP), An der Sternwarte 16, 14482 Potsdam, Germany \\
$^5$School of Astronomy and Space Science, University of Science and Technology of China, Hefei 230026, China \\
 }

\abstract{We report multi-wavelength observations of four solar
flares on 2014 July 07. We firstly select these flares according to
the soft X-ray (SXR) and extreme ultraviolet (EUV) emissions
recorded by the Extreme Ultraviolet Variability Experiment and
Geostationary Orbiting Environmental Satellites. Then their
locations and geometries are identified from the full-disk images
measured by the Atmospheric Imaging Assembly (AIA), and the
time delays among the light curves in different channels are
identified. The electron number densities are estimated using the
Differential Emission Measure method. We find that three of
four flares show strong emissions in SXR channels and high
temperature ($>$6~MK) EUV wavelengths during the impulsive phase,
i.e., AIA~131~{\AA} and 94~{\AA}, and then they emit peak radiation
subsequently in the middle temperature ($\sim$0.6$-$3~MK) EUV
channels. Moreover, they last for a long time and have smaller
electron densities, which are probably driven by the interaction of
hot diffuse flare loops. Only one flare emits radiation at
almost the same time in all the observed wavelengths, lasts for a
relatively short time, and has a larger electron density. It is also
accompanied by a type III radio burst. The bright emission
at the EUV channel could be corresponding to the associated erupting
filament. \keywords{Sun: flares -- Sun: radio radiation -- Sun: UV
radiation -- Sun: X-rays, gamma rays}}

\authorrunning{Dong Li et al.}            
\titlerunning{An investigation of flare emissions at multiple wavelengths}  
\maketitle

\section{Introduction}           
\label{intro}

Solar flare represents an impulsive and explosive release of
magnetic free energy by reconnection, which is always characterized
by a complex magnetic geometry
\citep{Masuda94,Shibata95,Priest02,Chen20}. The wavelength regime at
which flares radiate is very wide, ranging from radio through
optical and extreme ultraviolet (EUV) to soft/hard X-ray (SXR/HXR)
and even $\gamma$-rays beyond 1~GeV
\citep[e.g.,][]{Fletcher11,Benz17}. In a typical solar flare, a
large amount of magnetic free energy is released via magnetic
reconnection in the corona \citep{Shibata11,Li16,Xue16}, e.~g. as
described in the 2-D reconnection model \citep{Sturrock64} or the
CSHKP model \citep{Carmichael64,Sturrock66,Hirayama74,Kopp76}. In
this process, plasma will be heated to more than 10~MK in the
reconnection region, and electrons will be efficiently accelerated
to nonthermal energies. Subsequently, the released energy will be
transported away from the reconnection site in the form of
bi-directional outflows
\citep[e.g.,][]{Wangt07,Mann11,Liu13,Li19,Ning20}. A fraction of the
energy will be transported down to the transition region and
chromosphere or even photosphere along the newly formed flare loops
via thermal conduction and/or nonthermal electrons
\citep{Priest00,Battaglia09,Warmuth16,Emslie18}. During this
process, the flare loops can be clearly seen in SXR and EUV
wavelengths \citep{Sui03,Krucker08,Qiu09,Yan18}, and double
footpoints at the loop-legs and one or two loop-top sources can be
produced in HXR or microwave channels
\citep[e.g.,][]{Fletcher02,Lin03,Asai06,Chen17}, while the flare
ribbons are formed in H$\alpha$, visible, UV, and EUV wavebands
\citep{Temmer07,Wang14,Milligan15,Lit17,Song18}. Then the dense
material in the low solar atmosphere is rapidly heated to more than
10~MK and the resulting overpressure causes an upward bulk flow
along the flare loops at a fast speed, which is termed
`chromospheric evaporation' \citep{Fisher85,Brosius16,Li17,Tian18}.
At the same time, some of the released energy might travel up to the
outer corona or the interplanetary space in the form of nonthermal
electrons, which might produce type III radio bursts. Usually,
coronal mass ejections (CMEs) may also be accompanied by solar
flares \citep{Lin05,Zhang10,Cheng18}.

There are different classification schemes for solar flares. The
GOES class is a measurement of flare importance based on the peak
SXR~1$-$8~{\AA} flux measured by the Geostationary Orbiting
Environmental Satellites \citep[GOES,][]{Thomas71,Fletcher11}. At a
single wavelength, such as a SXR, HXR or radio channel, the temporal
evolution of the irradiance of a solar flare might be characterized
as either impulsive or gradual \citep{Pallavicini77,Ohki83}. This
has to be distinguished from the evolutionary phases of a solar
flare, which are defined according to specific characteristics of
the emission, i.e., pre-flare phase, impulsive phase, and decay
phase \citep{Hoyng81,Benz17}. Still another classification scheme
considerers the geometric properties of flares, which is an
important aspect since it reflects the underlying magnetic topology.
Accordingly, flares may display double or multiple ribbons, circular
ribbons, or might be very compact, which can be clearly seen in
H$\alpha$, UV or EUV images
\citep[see.,][]{Saint-Hilaire02,Su07,Radziszewski07,Zhang20}.
Conversely, multiple loops and footpoints are easily observed in
X-ray, EUV or radio images \citep{Masuda94,Su13,Chen17}. Solar
flares can also be considered as either eruptive or confined
\citep[e.g.,][]{Svestka92,Priest02,Ji03,Wang07,Shen11}. The eruptive
flares are often associated with the filament eruptions, CMEs,
and/or other ejecta \citep{Lin04,Forbes06,Cheng18}, while the latter
events are mostly caused by complex flare loop structures
\citep[e.g.,][]{Fan07,Guo10,Yang14,Zhang17}. Finally, a second peak
of warm coronal emissions (e.g., \fexvi~335~{\AA}) are found in
solar flares and named as the EUV late phase, which is separated
from the primary SXR peak by tens of minuets or even a few hours
\citep{Woods11}. Observations show that the EUV late phase flare
often shows double warm EUV peaks but only one SXR peak. The warm
EUV emission in the late phase is suggested to be originated from a
set of large-scale loops that are longer and higher than the primary
loops in the main phase of solar flare. It is thought to be related
to long-lasting cooling or additional heating \citep[e.g.,][and
references therein]{Dai13,Li14,Woods14,Liu15}.

Multi-wavelength observations of solar flares can provide an
opportunity to improve our understanding of the flare model. In this
paper, we performed a detailed analysis of four solar flares
occurring on 2014 July 07, which were simultaneously observed by the
GOES SXR channels, the Atmospheric Imaging Assembly
\citep[AIA,][]{Lemen12} and Extreme Ultraviolet Variability
Experiment \citep[EVE,][]{Woods12} aboard Solar Dynamics Observatory
(SDO) in SXR and EUV wavelengths, and the SWAVES instruments
\citep{Kaiser08} onboard Solar Terrestrial Relations Observatory
Behind (STEREO\_B). We mainly focused on the wavelength range over
which the flare was emitting and on their decay time in EUV
wavelengths and tried to search the causes of different emission
characteristics of solar flares at multiple wavelengths. This paper
is organized as following: the observations are introduced in
Section~\ref{Obs}, and the results are given in Section~\ref{Res},
then our conclusions and discussions are presented in
Section~\ref{Con}.

\section{Observations}
\label{Obs}

A few small solar flares occurred on the Sun between 11:45~UT and
14:40~UT on 2014 July 07. They were detected by the space-based
instruments in multi-wavelengths, whose details are listed in
table~\ref{tab}. During solar flares, SDO/AIA usually has two
different exposure times, i.e., one short exposure followed by one
long exposure. The long-exposure observations can increase the
signal-to-noise ratio, however, they can also result in image
saturation, in particular during a strong flare. Therefore, to avoid
the effect of image saturation, we here only use images with short
exposures, which means that the time cadence of AIA images in EUV
wavelengths is reduced to 24~s \citep{Lemen12,Ning17}.

\begin{table}
\caption{The details of the observational instruments in this paper}
\centering \setlength{\tabcolsep}{10pt}
\begin{tabular}{c c c c c c c}
 \hline\hline
Instruments   &  Channels             &   Cadence (s)     &  Descriptions        \\
\hline
              &  1$-$8~{\AA}          &   $\sim$2.0      &    SXR               \\
GOES          &  0.5$-$4~{\AA}        &   $\sim$2.0      &    SXR               \\
\hline
SWAVES        &  0.023$-$16.025~MHz   &    60.0           &    radio             \\
\hline
SDO/AIA       &  94~{\AA}$-$335~{\AA}  &    24.0           &    EUV               \\
\hline
             &  1$-$70~{\AA}          &    0.25           &    SXR               \\
SDO/EVE/ESP  &   180~{\AA}            &    0.25           &   172$-$206~{\AA}    \\
             &   300~{\AA}            &    0.25           &   280$-$316~{\AA}    \\
 \hline\hline
\end{tabular}
\label{tab}
\end{table}

\begin{figure}[h!!!]
\centering
\includegraphics[width=\linewidth,clip=]{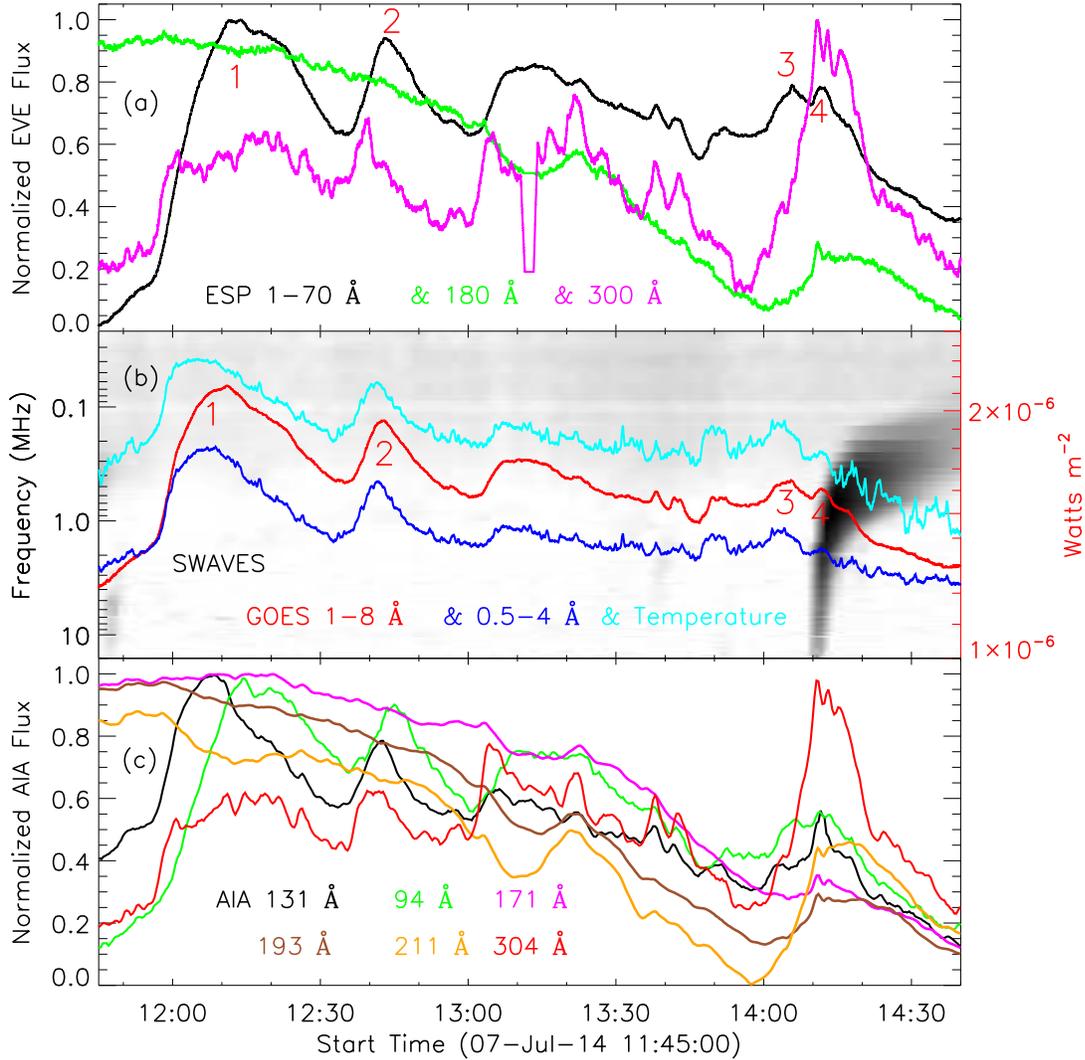}
\caption{Panel~(a): Normalized SDO/EVE fluxes between 11:45~UT and
14:40~UT on 2014 July 07 in ESP~1$-$70~{\AA} (black), 180~{\AA}
(green), and 300~{\AA} (magenta). Panel~(b): GOES fluxes in SXR
1$-$8~{\AA} (red) and 0.5$-$4~{\AA} (blue), and temperature (cyan).
Panel~(c): Normalized light curves integrated over the full-disk Sun
in AIA~131~{\AA} (black), 94~{\AA} (green), 171~{\AA} (magenta),
193~{\AA} (brown), 211~{\AA} (orange), and 304~{\AA} (red). The
context image in panel~(b) is radio spectrogram derived from SWAVES.
\label{spect}}
\end{figure}

Figure~\ref{spect} shows the normalized light curves integrated over
the whole Sun at multiple wavelengths detected by the SDO/EVE, GOES,
and SDO/AIA, respectively. Here, the light curves measured
by the EVE and AIA are normalized by the
equation$\frac{I-I_{min}}{I_{max}-I_{min}}$, where I is the observed
intensities, $I_{min}$ and $I_{max}$ represent the minimum and
maximum intensities, respectively. The light curves in panel~(a)
were obtained from the EUV SpectroPhotometer
\citep[ESP,][]{Didkovsky12} onboard SDO/EVE. It can be seen that the
SXR light curve in ESP~1$-$70~{\AA} exhibits several peaks, such as
peaks~1$-$4. Then the similar peaks can be found in the EUV
flux at ESP~300~{\AA}, which is a bit earlier than the SXR peaks.
Moreover, the first three peaks (1$-3$) at ESP~300~{\AA} is much
smaller than the last one (4), which is the weakest peak at
ESP~1$-$70~{\AA}. On the other hand, only the last peak could be
easily identified in the EUV irradiance at ESP~180~{\AA}. The
ESP~180~{\AA} channel contains line emissions that mainly come from
the plasmas around 1$-$2~MK, while the ESP~260~{\AA} and 300~{\AA}
channels are dominated by \heii\ lines \citep{Didkovsky12,Woods12}.

Figure~\ref{spect}~(b) presents the GOES flux in SXR~1$-$8~{\AA}
(red) and 0.5$-$4~{\AA} (blue), from which, we can find the similar
peaks to the ESP SXR light curve, as indicated by 1$-$4. Thus, those
four peaks can be identified as the C-class flare. Moreover, they
exhibit impulsive peaks in the temperature profile derived from the
GOES fluxes using the isothermal assumption \citep{White05}, which
are consistent with the SXR flare peaks, indicating these four
flares are hot. Therefore, these four flares display almost the same
evolutionary behavior in the SXR flux recorded by the GOES and
EVE/ESP. Finally, the radio spectrogram shows a strong and
well-developed type III burst at frequencies between $\sim$0.1~MHz
and $\sim$16~MHz during the flare~`4', as shown in the context image
measured by the SWAVES.

To confirm that the EUV flux recorded by the EVE/ESP is credible, we
then plot the SDO/AIA light curves integrated over the full-disk Sun
in EUV passbands, as shown in Figure~\ref{spect}~(c). The
SDO/AIA fluxes in high ($>$6~MK) temperature channels exhibit the
similar flare peaks as the SXR light curves, i.e., AIA~131~{\AA}
($\sim$10~MK), and 94~{\AA} ($\sim$6.3~MK), and the light curve in
low ($<$0.1~MK) temperature channel also shows similar peaks with
the ESP~300~{\AA} flux, such as AIA 304~{\AA} ($\sim$0.05~MK). The
time delays among these light curves can be clearly seen in the
first three flares, but not found in the last flare. On the other
hand, the SDO/AIA fluxes in middle temperature
($\sim$0.6$-$2~MK) wavelengths such as AIA~211~{\AA} ($\sim$2.0~MK),
193~{\AA} ($\sim$1.6~MK), and 171~{\AA} ($\sim$0.6~MK) only display
the pronounced peak for the flare~4, which are almost simultaneous.
This observational result is consistent with EVE/ESP observations at
different channels.

\section{Results}
\label{Res}
\begin{figure}[h!!!]
\centering
\includegraphics[width=\linewidth,clip=]{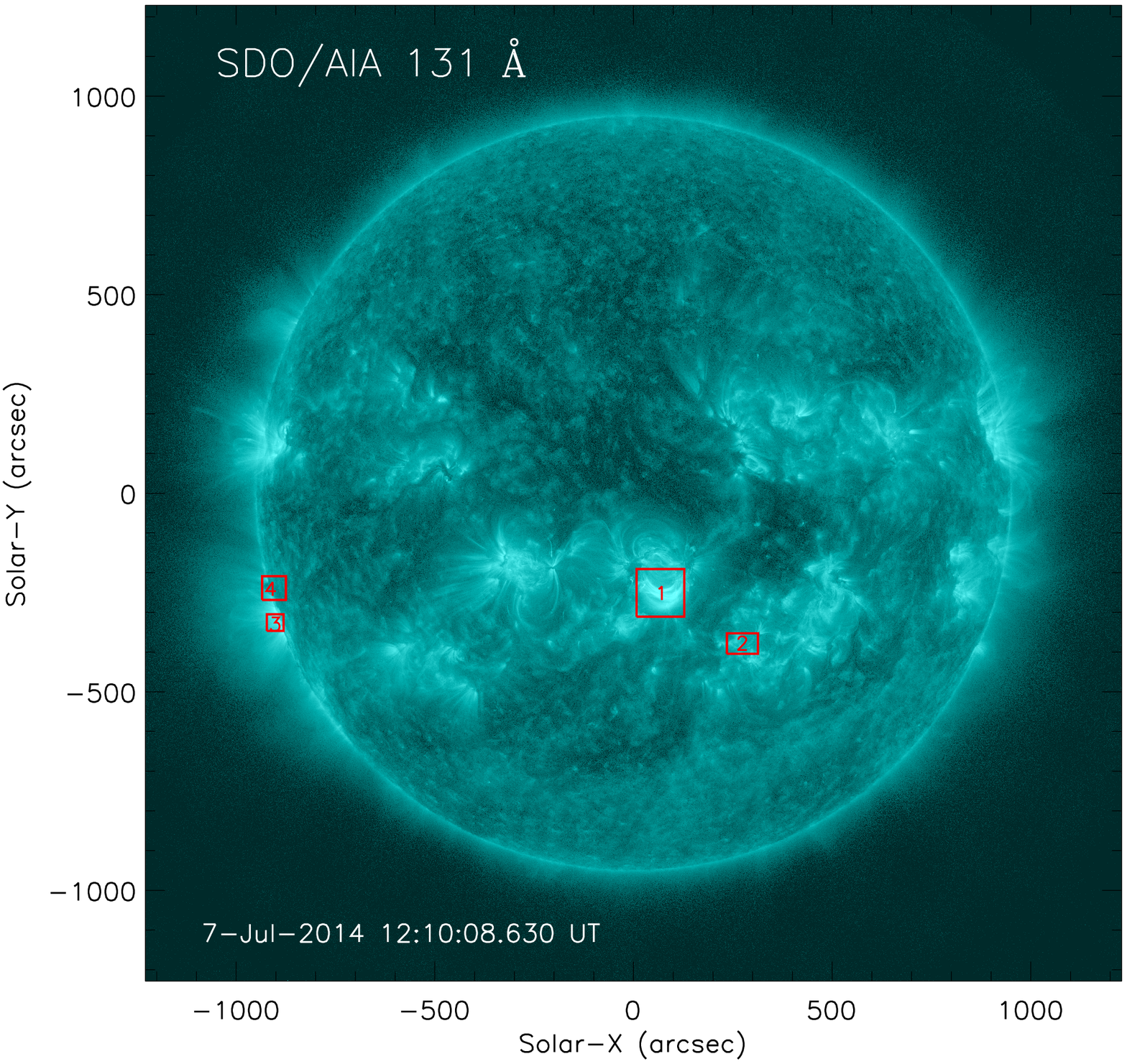}
\caption{Snapshot of the full Sun in AIA~131~{\AA} at 12:10:08~UT on
2014 July 07. The red rectangles indicate the flare subregions in
this study. \label{snap}}
\end{figure}

\begin{figure}[h!!!]
\centering
\includegraphics[width=\linewidth,clip=]{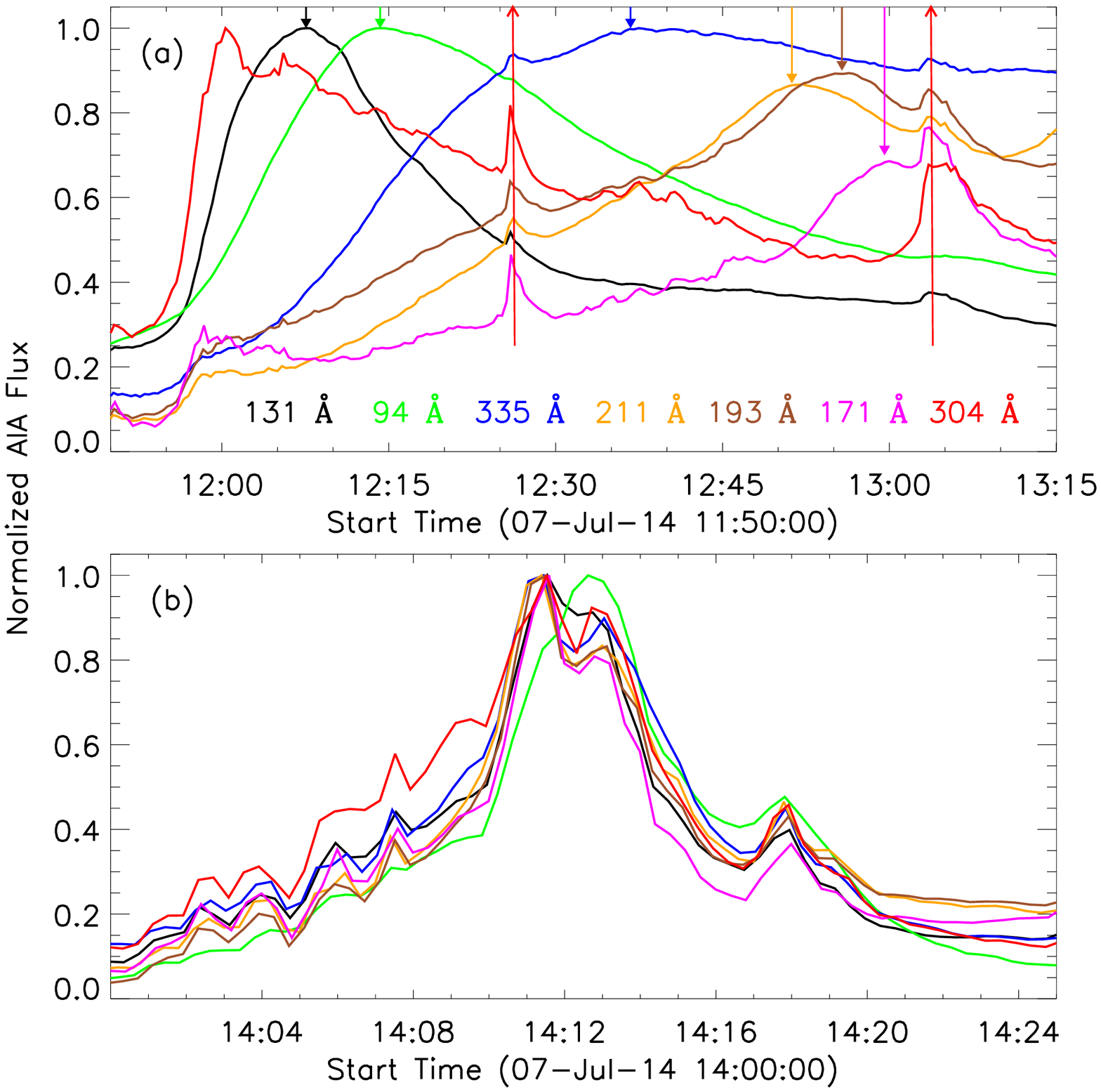}
\caption{Normalized EUV fluxes for flares 1 and 4 at wavelengths of
AIA~131~{\AA} (black), 94~{\AA} (green), 335~{\AA} (blue), 211~{\AA}
(orange), 193~{\AA} (brown), 171~{\AA} (magenta), and 304~{\AA}
(red) derived from the flare subregions outlined by the red
rectangles in Figure~\ref{snap}. The red arrows indicate the small
jets, while the other color arrows marks the peak time at
multiple wavelengths. \label{flux}}
\end{figure}

For further analysis, four flare subregions, which are the source
regions of the solar flares under study, are selected. These
subregions are marked with the red rectangles in Figure~\ref{snap},
and each of them is assigned a number. Two of the flare regions
numbered by `1' and `2' are located near the solar disk center,
while the other two (`3' and `4') are located at the solar limb.
Figure~\ref{flux} shows the normalized fluxes resulting from an
integration of the SDO/AIA images over the subregions for flares~1
and 4. As can be seen in panel~(a), the flare~1 exhibits a
pronounced peak in both high and low temperature EUV channels, and
the peak time of AIA~304~{\AA} is a little earlier than that of
AIA~131~{\AA} and 94~{\AA}. Moreover, they decay slowly and can
maintain a long time. This flare shows a weak peak during the
impulsive phase of solar flare in the middle temperature EUV
channels, such as AIA~335~{\AA}, 211~{\AA}, 193~{\AA} and~171~{\AA},
which is almost at the same time with that in AIA~304~{\AA}. On the
other hand, the strong peak in the middle temperature EUV channels
is seen during the flare decay phase, which could be considered as
the post flare loops, as can be seen in the movie of flare1.mp4. The
strong peaks both in high and middle temperature EUV channels appear
subsequently, as indicated by the solid color arrows. All these
observational facts suggest a slowly cooling precess in this flare.
Panel~(b) shows that the flare~4 emits strongly radiation at almost
all the SDO/AIA wavebands, including the high, middle and low
temperature channels. Moreover, the flare emission in
different channels reach the peaks at almost the same time. Then,
it decays quickly and has a short lifetime in all EUV wavelengths.
In a word, it releases energy drastically and quickly in all the
observed wavelengths. In addition, the flare peaks in
flares~2 and 3 also appear subsequently in different EUV
wavelengths, as shown in Figure~\ref{flux1}, which are similar to
flare~1.

\begin{figure}[h!!!]
\centering
\includegraphics[width=\linewidth,clip=]{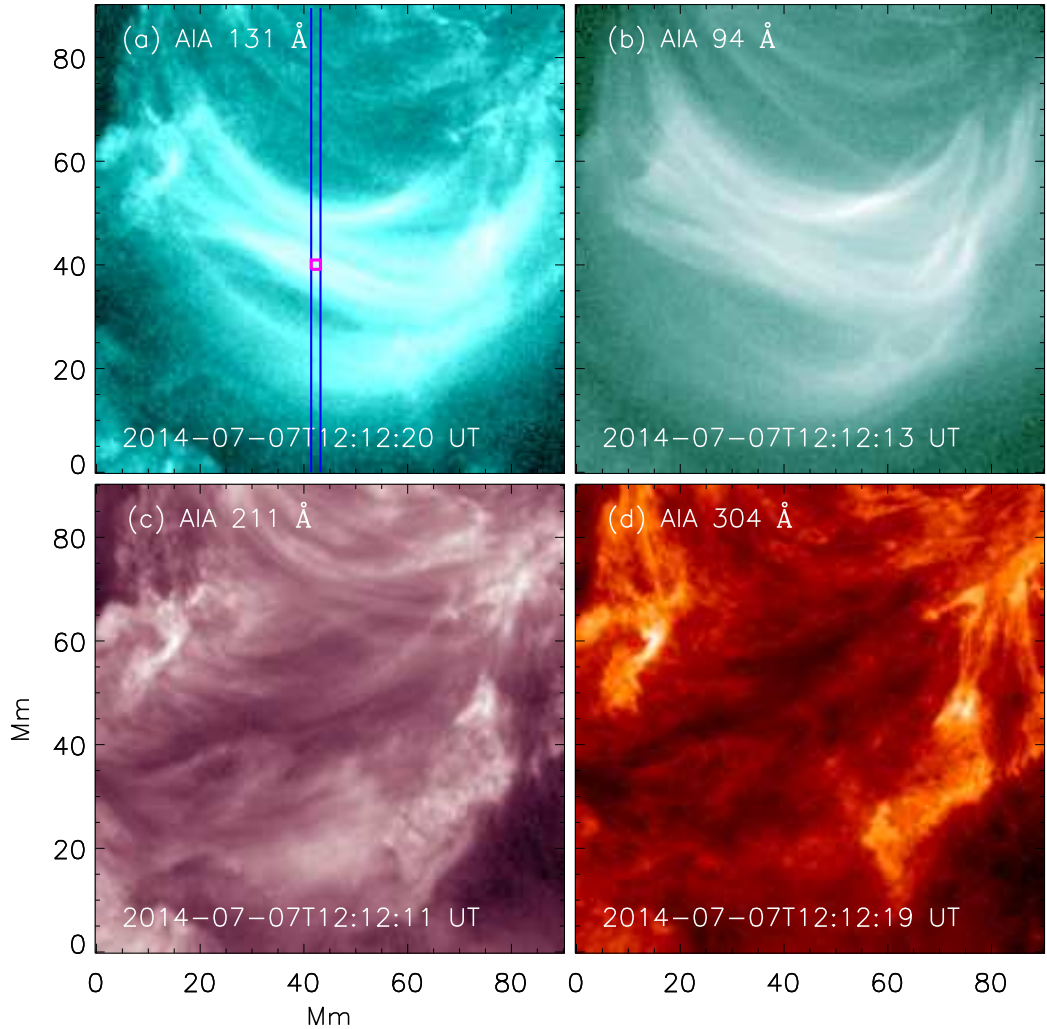}
\caption{Small FOV images for flare~1 at around 12:12~UT in AIA
131~{\AA} (a), and 94~{\AA} (b), 211~{\AA} (c) and 304~{\AA} (d).
The magenta box outlines the flare loop region
(2.4\arcsec$\times$2.4\arcsec) used to perform the DEM analysis, and
double blue lines outline the positions used to integrate the
intensities along the north-south direction. \label{img1}}
\end{figure}

\begin{figure}[h!!!]
\centering
\includegraphics[width=\linewidth,clip=]{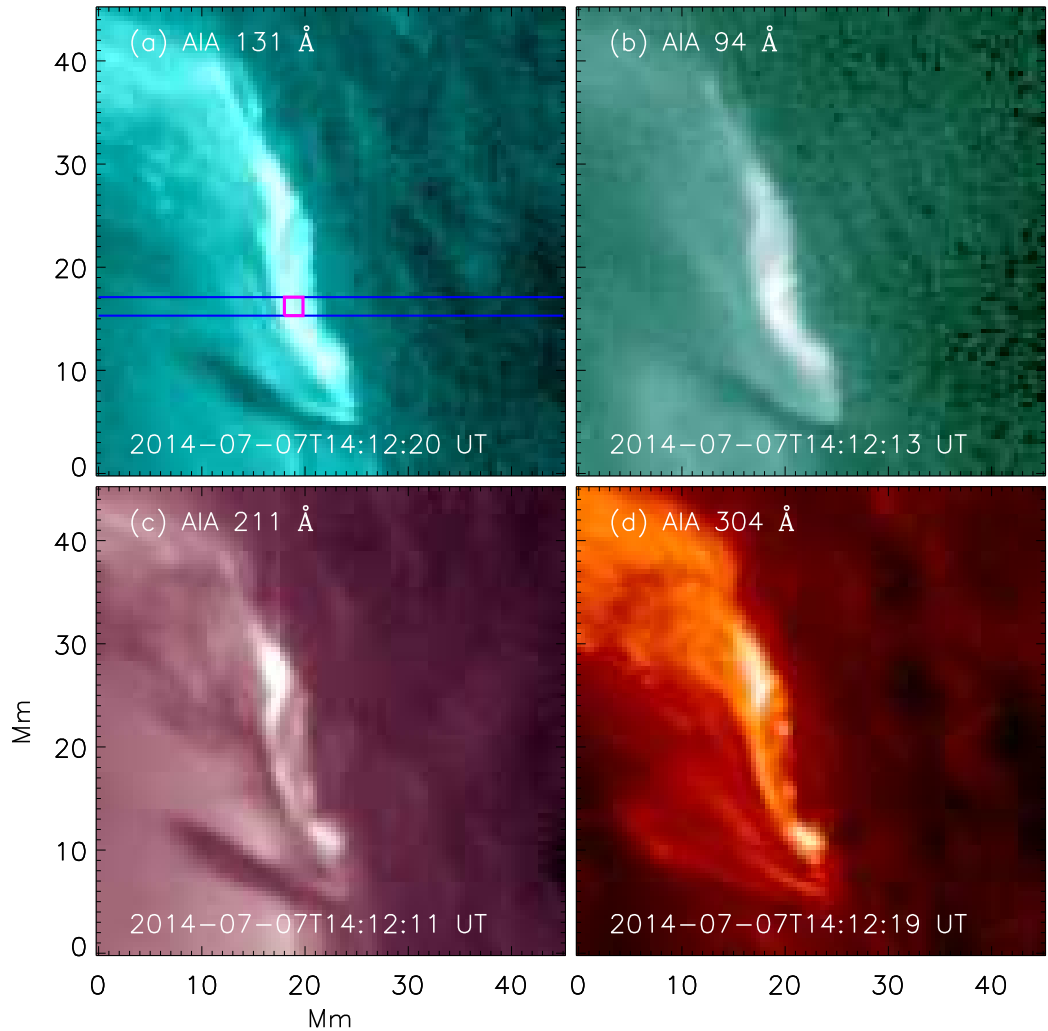}
\caption{Same as Figure~\ref{img1}, but for flare~4 at around
14:12~UT. \label{img4}}
\end{figure}

To look closely the geometries of these flares, we then plot the AIA
images with a small field-of-view (FOV) during flare peak times.
Figure~\ref{img1} shows SDO/AIA images in EUV wavelengths for the
flare~1 with a FOV of about 90$\times$90~Mm$^2$, as outlined by the
red box~`1' in Figure~\ref{snap}. Panel~(a) and (b) presents the
small FOV images in AIA~131~{\AA} and 94~{\AA} at a time about
12:12~UT, which is around the peak time of flare~1 in
GOES~1$-$8~{\AA} (Figure~\ref{spect}) and AIA~94~{\AA}
(Figure~\ref{flux}). We can clearly see a broad bundle of flare
loops in these two high temperature EUV wavelengths. However, it is
a combination of the limited spatial resolution of AIA and the
simple fact that multiple optically thin loops overlayed on each
other just cannot be easily separated. Therefore, we consider them
together and use a Gaussian function to fit them, which are
integrated along two blue lines in panel~(a). Thus, we estimate a
loop thickness of about 13.6~Mm, which is regarded as the full width
at half maximum (FWHM). Then, panels~(c) and (d) display the same
FOV images in AIA~211~{\AA} and 304~{\AA} at almost the same time
i.e., around 12:12~UT. From which, we can see two weak flare
ribbons, implying that the flare emission in these two EUV
wavelengths mainly appears in the ribbons during the impulsive
phase. The flare loops, on the other hand, could not be seen in
these middle temperature AIA wavelengths at impulsive and main
phases that close to the flare peak time. All these AIA images
suggests that the diffuse flare loops in the high temperature
channels connect double ribbons seen in middle and low temperature
wavelengths. The movie of flare1.mp4 suggests that the flare~1 is
constituted of strong hot loops but weak ribbons at impulsive and
main phases. Finally, at the decay phase that far away from the
flare peak time, some post flare loops \citep{Li09a,Li09b} can be
found subsequently. We would like to point out that these
post flare loops appear successively, suggesting a cooling process
during this flare. We also notice that some small peaks appear at
around 12:26~UT and 13:02~UT (red arrows) at all the EUV channels in
Figure~\ref{flux}~(a), which are from two small-scale jets, as shown
in the movie of flare1.mp4.

For flare~4 which is located at the solar limb, Figure~\ref{img4}
shows the AIA images at a time near 14:12~UT with a small FOV of
around 45$\times$45~Mm$^2$, as marked by the red box~`4' in
Figure~\ref{snap}. These images and the movie of flare4.mp4 show
that flare~4 is very compact and emits radiation in all the EUV
wavelengths. This flare is associated with a small-scale filament
eruption, as can be seen in the movie. It also causes a solar jet
and leads to a weak CME \citep[see detail in][]{Lu19}, as well as a
type III radio burst (Figure~\ref{spect}~b). Using the same
Gaussian fit method for the intensities integrated along
two blue lines in panel~(a), we then estimate the loop thickness
(i.e., FWHM) of $\sim$2.2~Mm. Figures~\ref{img2} and \ref{img3} show
the AIA images at around 12:44~UT and 14:01~UT for flares~2 and 3,
respectively. The subregions considered here have sizes of
58$\times$39~Mm$^2$ and 32$\times$32~Mm$^2$, respectively. These
images present a similar analysis for flares~2 and 3, both of which
show a consistent result with the flare~1. In other words, they are
constituted of bright hot loops but faint ribbons with low
temperature at impulsive and main phases that close to the flare
peak time. In the decay phase, the flare loops appear
successively at different EUV channels. Note that this
characteristic is independent of the flare location on the disk, so
it can not be explained by projection effects.

\begin{figure}[h!!!]
\centering
\includegraphics[width=\linewidth,clip=]{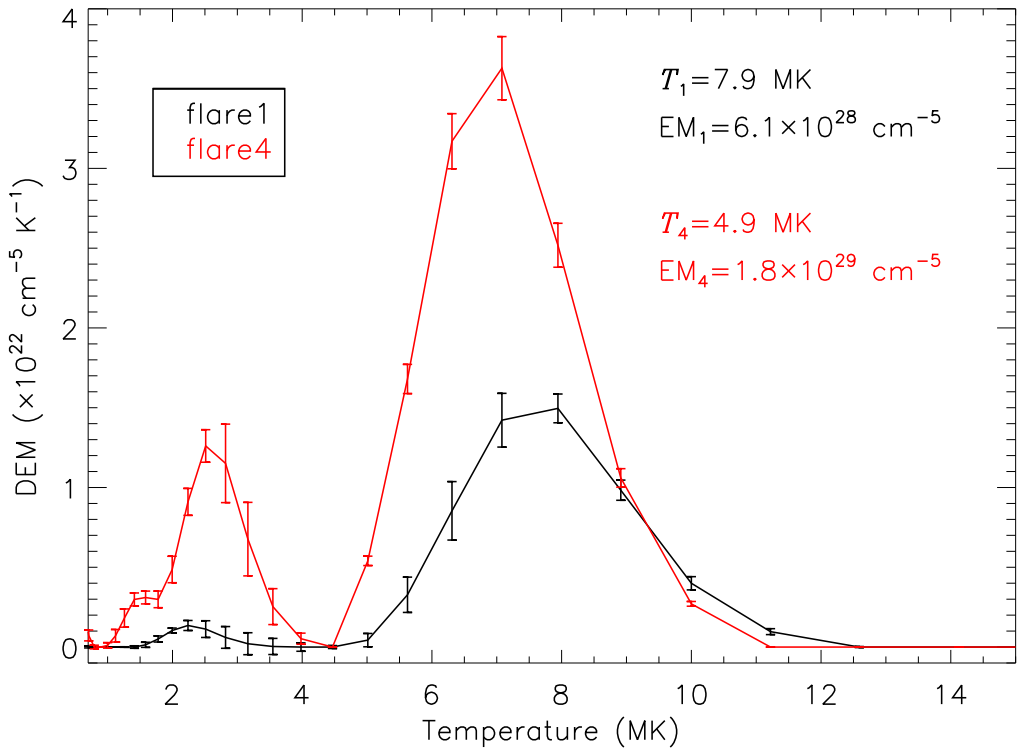}
\caption{DEM curves at the selected positions (magenta box) for
flare~1 (black) and flare~4 (red). The error bars are estimated from
the 100 MC simulations. The EM-weighted mean temperature and total
EM are given. \label{dem}}
\end{figure}

In order to determine the electron number density ($n$) in the flare
regions, we perform a differential emission measure (DEM) analysis
using observations made in six AIA EUV wavelengths. We employ the
sparse inversion code by \cite{Cheung15} in a recently modified
version \citep{Su18}, and the DEM uncertainties are estimated from
100 Monte Carlo (MC) simulations, i.e., their standard deviation.
Assuming optically thin emissions in solar flares, the total
emission measure (EM) and the EM-weighted mean temperature ($T$) in
the selected flare regions can be derived from the DEM results.
Figure~\ref{dem} presents the DEM curves at the selected positions
(magenta box), where flare emissions in high temperature channels
are strongest. It can be seen that the flare~1 (black) shows a
strong peak at the higher temperature of around 8~MK, suggesting hot
loops in flare~1 during the impulsive phase. On the other
hand, the flare~4 (red) displays double peaks at the higher
($\sim$7~MK) and lower ($\sim$2.5~MK) temperatures, implying
multi-thermal structures in this flare. Then, the electron number
density is calculated according to $n = \sqrt{\frac{EM}{w}}$
\citep[see,][]{Aschwanden05}, where $w$ represents the effective
line-of-sight (LOS) depth, which is adopted as the loop thickness
here. We can estimate the EM-weighted mean temperature and electron
number density at a small region (magenta box) for flares~1 and 4,
such as $T_1=7.9$~MK, $T_4=4.9$~MK, $n_1=6.7\times10^9$~cm$^{-3}$,
$n_4=2.9\times10^{10}$~cm$^{-3}$, respectively. It can be seen that
flare~4 has a larger number density but a lower temperature compared
to flare~1, indicating that flare~4 is denser but cooler than the
first flare.

\section{Conclusion and Discussion}
\label{Con}

Using multiple-wavelength observations measured by SDO/EVE, SDO/AIA,
GOES, and SWAVES, four solar flares occurring on 2014 July 07 were
studied. They all emitted thermal radiation generated by hot plasma
in SXR wavelengths, i.e, ESP~1$-$70~{\AA}, GOES~1$-$8~{\AA}, and
0.5$-$4~{\AA}. The first three flares (flares~1$-$3) showed strong
emissions in SXR and high temperature EUV wavelengths during the
impulsive phase, and no radio burst was detected. Moreover, they
could maintain a long lifetime. During their decay phases that far
away from the flare peak time, some post flare loops appear
subsequently at the middle temperature channels, such as
AIA~335~{\AA}, 211~{\AA}, 193~{\AA} and 171~{\AA}, suggesting long
cooling time from high to low temperatures. Conversely, the last
flare (flare~4) in our samples was emitting simultaneously in almost
all the SXR and EUV channels and was accompanied by a strong type
III radio burst, and it was fade quite quickly. We stress
that the bright emissions in EUV wavelengths could reach peaks at
almost the same time, which is most likely to be an erupting
filament.

The reason why the first three flares are different from the last
one is most likely associated with their geometric configurations
\citep{Zhang17,Sarkar18}. The first three flares show very
diffuse loops in high temperature channels at beginning, and then
appear subsequently in other EUV wavebands, while the flare~4
exhibits a compact feature in all EUV wavelengths. For the first
three flares, magnetic reconnection might be triggered by the
interactions of hot flare loops at beginning
\citep[see,][]{Torok05,Guo10,Liu14}. In these events, the
lack of emission from low/middle temperature plasmas is found during
the impulsive phase, i.e., before the peak of SXR flux. However, the
appearance of flare ribbons and post flare loops in the middle
temperature channels support the energy transport from the
reconnection region to the chromosphere and the occurrence of
chromospheric evaporation during these flares, which is similar to
particle acceleration with subsequent chromospheric evaporation
\citep[e.g.,][]{Fisher85,Milligan15,Brosius16,Tian18}. On the other
hand, the lower number density and longer flare loops should imply
smaller radiative and conductive losses, which would result in a
significantly longer cooling time as compared to the last event
\citep{Cargill95,Aschwanden01,Benz17}. The bulk of the released
energy will thus be retained in flare loops for a long duration,
while comparatively little energy will be slowly transported to the
deeper atmospheric layers. This is also consistent with the
subsequent post flare loops at middle temperature channels, which
are far away from the flare peak time. Conversely, flare~4 has high
density, resulting in larger radiative losses -- a possible
mechanism for energy transport to deeper layers. This is confirmed
by the DEM results during the flare peak time, which show
hot temperatures but low densities in flare~1, and a
multi-thermal temperature but high density in flare~4.

We would like to point out that the first three flares might
be caused by the reconnection between complex hot flare loops, so
there is no open field line generated during the flare processes.
Thus, there is no type III radio burst during these three flares,
which is also the difference of the first three flares from the last
flare associated with the filament eruption. On the other hand, the
flare~4 is associated with an erupting filament, the accelerated
electron beams then propagate upward along the open magnetic field
lines, generating the type III radio burst. Unfortunately, no
appropriate HXR observations were available for our event sample, so
we cannot make a definite conclusion about the role of energetic
electrons. To make this conclusion more robust, we will investigate
more cases which will include explicit HXR observations in future,
since they provide direct evidences of nonthermal electrons in the
low solar atmosphere \citep{Fletcher11,Benz17}. The Hard X-ray
Imager \citep{Su19,Zhang19} on board Advanced Space-based Solar
Observatory \citep{Gan19,Huang19} and the Spectrometer/Telescope for
Imaging X-rays \citep{Krucker20} aboard Solar Orbiter
\citep{Muller13} could help to solve this problem, in particular for
their joint observations in HXR channels \citep[e.g.,][]{Krucker19}.

It is worthwhile to stress that the first three flares are
geometrically quite extended, $-$ in particular compared to flare~4.
Taking the flare~1 for example, the double footpoints connected by
flare loops are separated by $s~\sim$~70~Mm, which corresponds to
some of the largest values of the RHESSI flares found by
\cite{Warmuth13}. Considering optically thin emissions in solar
flares, the broad bundle of flare loops should correspond to a large
LOS depth ($d$), thus resulting in low densities and therefore a low
radiative cooling rate. Conversely, when the loop half-length
($L\approx\pi~s/4$) is adopted at the thermal gradient length, a
comparatively low conductive cooling rate would result. Using the
key parameters derived from flare~1, such as the number density
($n_1$), the plasma temperature ($T_1$), and loop half-length ($L$),
the conductive cooling($\tau_{con}$) and radiative cooling
($\tau_{r}$) timescales \citep[cf.][]{Cargill04,Reale14} could be
estimated, which are $\tau_{con}\sim$~29~minutes and
$\tau_{r}\sim$~166~minutes, respectively. Similarly, the conductive
cooling ($\tau_{con}'$) and radiative cooling ($\tau_{r}'$)
timescales of flare~4 can also be estimated from its key parameters,
such as $\tau_{con}'\sim$~44~minutes and $\tau_{r}'\sim$~19~minutes.
It should be mentioned that the role of conduction is still under
discussion. There are theoretical arguments that suggest conduction
might be significantly suppressed in solar flares \citep[e.g.,][and
references therein]{Bian18,Emslie18}. Conversely, there is evidence
for conductive evaporation in some events
\citep[e.g.,][]{Battaglia09,Zhang13}. In general, shorter flare
loops should lead to smaller conductive timescales
\citep{Aschwanden01}, but in our case (flare~4) this is
overcompensated by the multi-thermal temperature. The conductive
timescales are comparable in our flares, so conduction is not
favored for the explanation of the vastly different decay
timescales. In other words, our interpretation does not depend on
the exact role of conduction: the vastly different radiative cooling
timescales are sufficient for explaining the different temporal
evolution of the compact flare 4 as opposed to the extended
flares~1$-$3.

It is also necessary to stress that the long lifetime of the first
three flares is essentially distinct from the EUV late phase of
solar flares discovered by SDO/EVE and SDO/AIA, which refers to the
second peak in warm EUV (i.e., \fexvi~335~{\AA}) channels
\citep{Woods11,Woods14}. Since the first three flares in our work do
not exhibit the second large peak which is separated from the
impulsive peak in high temperature EUV wavelengths, such as AIA
131~{\AA} and 94~{\AA}. Moreover, the middle temperature EUV (e.g.,
AIA 335~{\AA}, 211~{\AA}, 193~{\AA} and 171~{\AA}) channels show
peak successively in our flares, as can be seen in
Figure~\ref{flux}.

\normalem
\begin{acknowledgements}
We acknowledge the anonymous referee for his/her valuable comments.
We thank the teams of SDO/EVE, SDO/AIA, GOES, and STEREO for their
open data use policy. This work is supported by NSFC under grants
11973092, 11790300,11790302, 11729301, 11603077, and the Strategic Priority
Research Program on Space Science, CAS, Grant No. XDA15052200 and
XDA15320301. D.~Li is also supported by the Specialized Research
Fund for State Key Laboratories and CAS Strategic Pioneer Program on
Space Science (KLSA202003). The Laboratory No. 2010DP173032. The
work of A. W. was supported by DLR under grant No. 50 QL 1701.
\end{acknowledgements}

\label{lastpage}

\clearpage
\appendix
\section{AIA fluxes for flares 2 and 3}
\begin{figure}[h!!!]
\centering
\includegraphics[width=\linewidth,clip=]{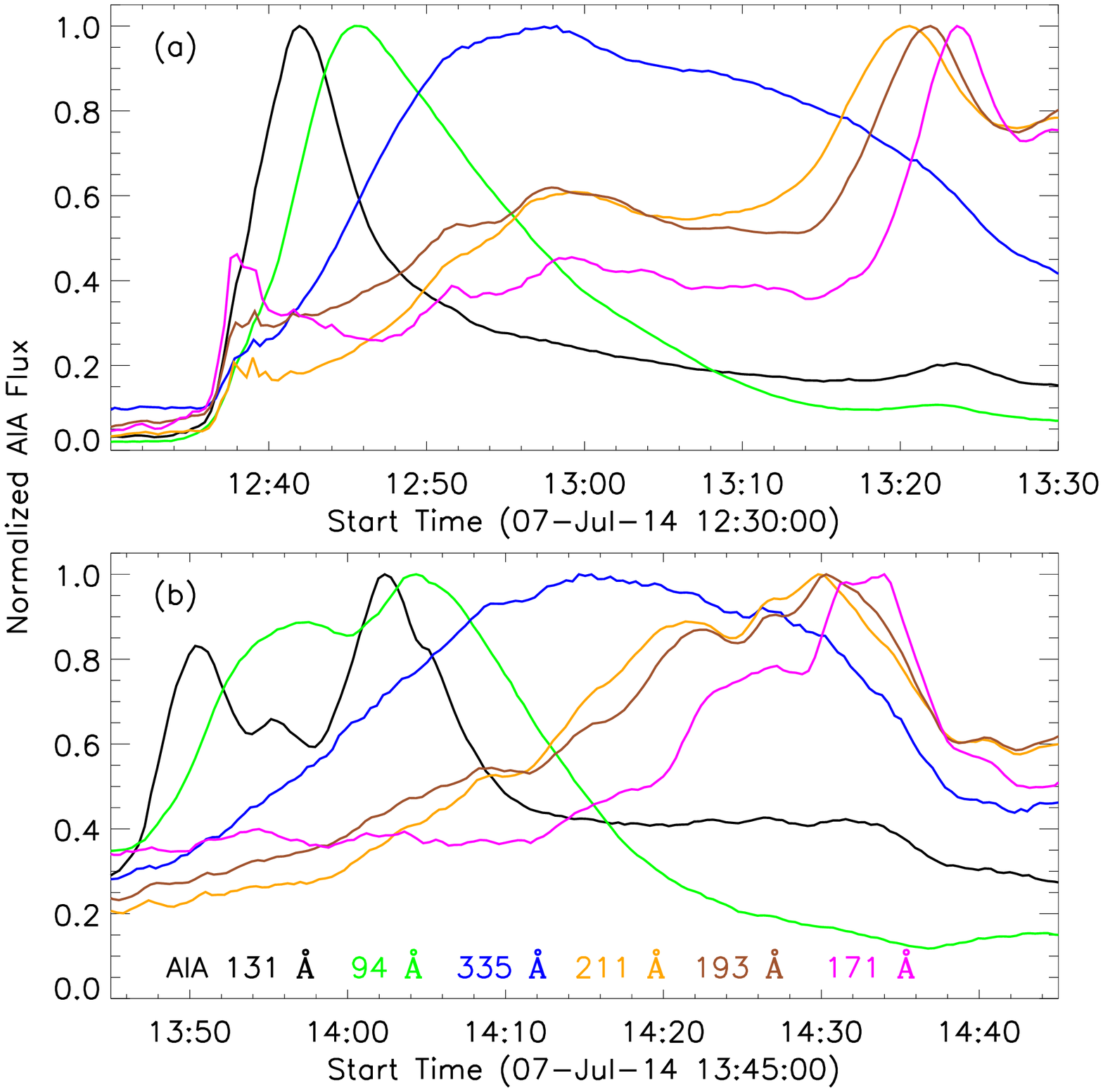}
\caption{Normalized EUV fluxes for flares 2 and 3 at wavelengths of
AIA 131~{\AA} (black), 94~{\AA} (green), 335~{\AA} (blue), 211~{\AA}
(orange), 193~{\AA} (brown), and 171~{\AA} (magenta), which are
derived from the flare subregions outlined by the red rectangles in
Figure~\ref{snap}. \label{flux1}}
\end{figure}

\section{AIA images for flares 2 and 3}
\begin{figure}[h!!!]
\centering
\includegraphics[width=\linewidth,clip=]{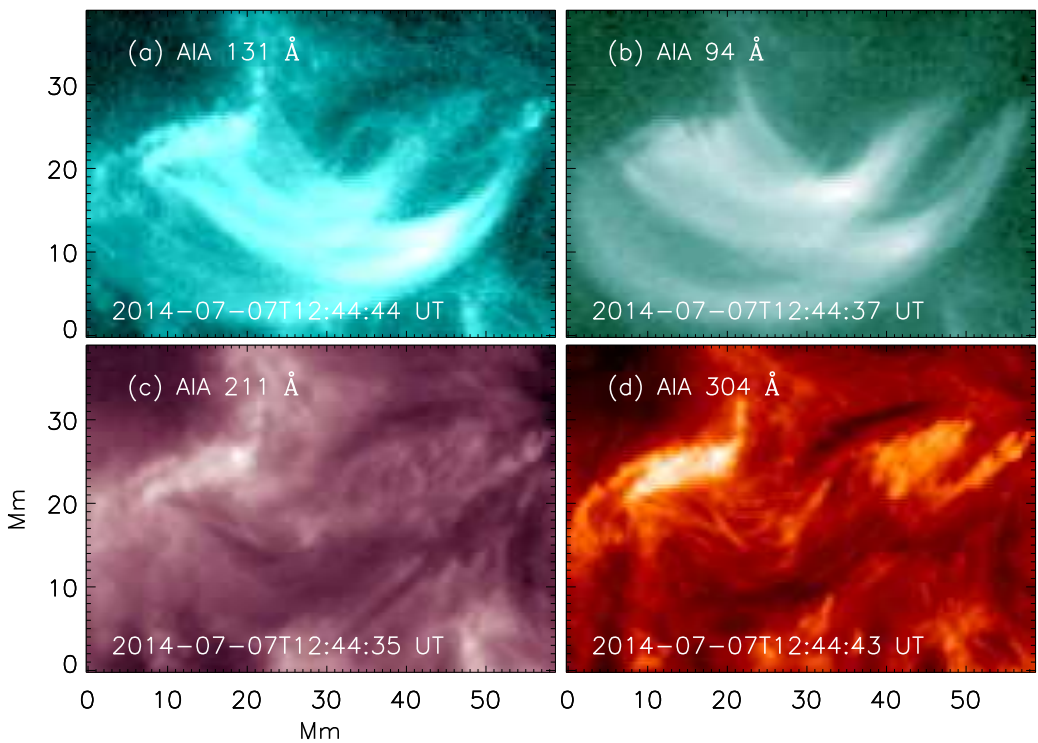}
\caption{Small FOV images for flare~2 at around 12:44~UT in AIA
131~{\AA} (a), 94~{\AA} (b), 211~{\AA} (c) and 304~{\AA} (d).
\label{img2}}
\end{figure}

\begin{figure}[h!!!]
\centering
\includegraphics[width=\linewidth,clip=]{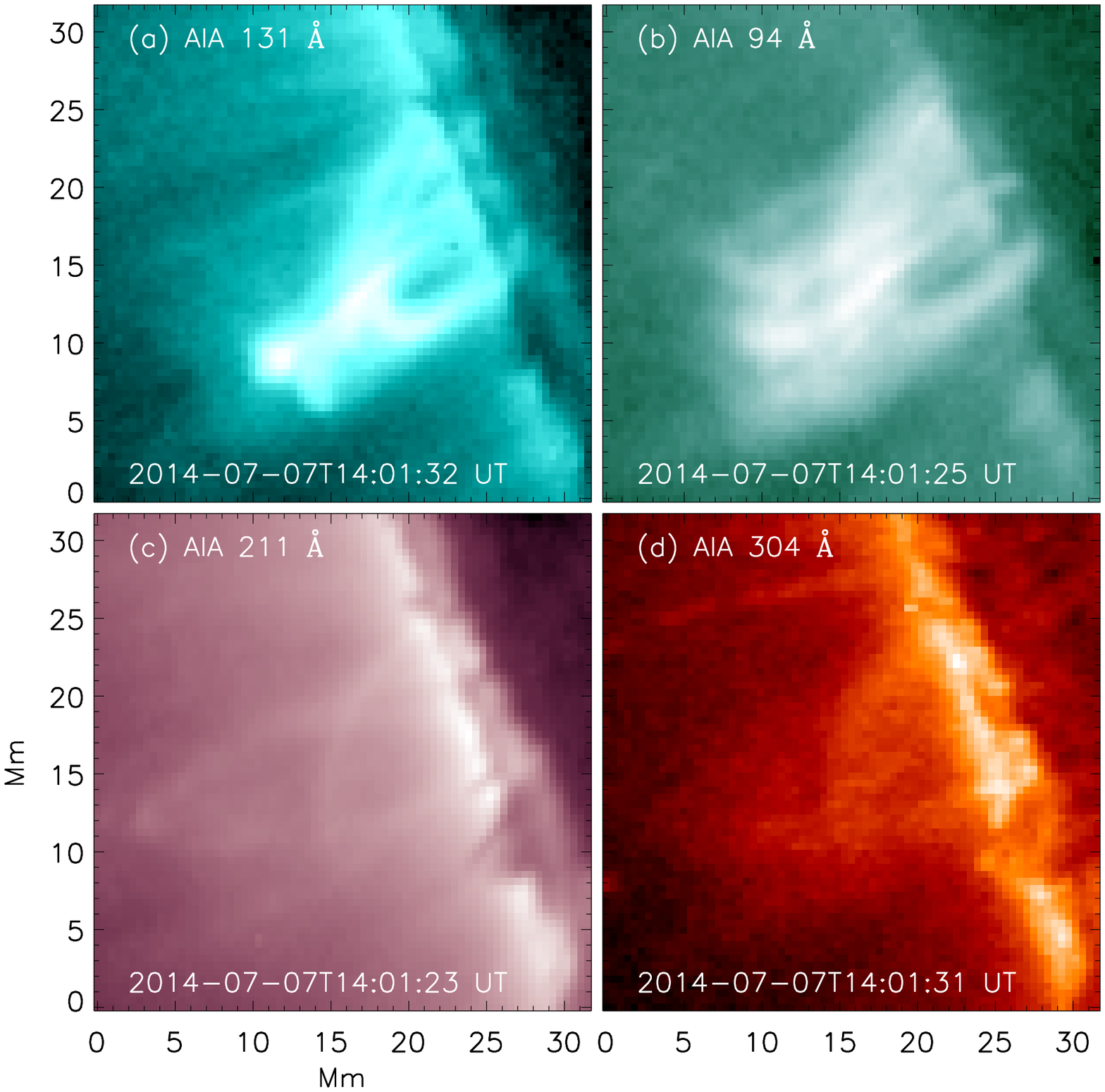}
\caption{Same as Figure~\ref{img2}, but for flare~3 at around
14:01~UT. \label{img3}}
\end{figure}

\end{document}